# Experimental evidence of seismic ruptures initiated by aseismic slip

Yohann Faure[a], Elsa Bayart[a,*]

[a]*ENS de Lyon, CNRS, Laboratoire de physique, F-69342 Lyon, France*

**Abstract**

Seismic faults release the stress accumulated during tectonic movement through rapid ruptures or slow slip events. The slow slip events play a crucial role in the seismic cycle as they impact the occurrence of earthquakes. However, the mechanisms by which a slow-slip region affects the dynamics of frictionally locked regions remain elusive. Here, building on model laboratory experiments, we establish that a slow-slip region acts as a nucleation center for seismic rupture, thereby enhancing earthquakes' frequency. We emulate slow-slip regions by introducing a granular material patch along part of a laboratory fault. By measuring the response of the fault to shear, we show that the role of the heterogeneity is to serve as a seed crack for rapid ruptures, reducing fault shear resistance. Additionally, by varying the external normal load, we show that the slow-slip region extends beyond the heterogeneity, demonstrating that fault composition is not the only requirement for slow-slip, but that load also plays a role. Our findings demonstrate that fracture concepts single out the very origin of earthquake nucleation and slip dynamics in seismic faults. The interplay between slowly-slipping and locked regions that we identify provides a promising avenue to monitor fault propagation and mitigate seismic hazards.

Seismic faults release accumulated stresses via both rapid ruptures, called seismic events, and slow slip events (SSEs)[1]. SSEs were first discovered in subduction zones, but improved measurement techniques have shown that they also occur within subsurface strike-slip faults[2]. SSEs can affect large areas of faults, or can be confined to localized asperities. These very small events are detected indirectly by the seismic events they induce, also known as tremors[3-5]. While their existence is well-established, the role of SSEs in the seismic cycle is still unclear. In some cases, they can be responsible for triggering large earthquakes[6], while in other cases they occur periodically without being earthquake precursors[7]. Understanding the interaction mechanisms between uncoupled slipping zones within a fault or fault system, and coupled frictionally locked zones is essential, especially for the design of fault monitoring strategies[8]. A critical question is what fault properties are required for SSEs to occur. Slow slip areas are commonly modeled as velocity strengthening zones[9] that cannot rupture seismically. However, simple geometric complexity with no assumptions about the frictional properties of the interface is sufficient to induce slow slip[10]. In addition, observations have shown that seismically slipping zones can exhibit slow slip[11-13]. This finding raises the question of the importance of the boundaries of slow slip zones for seismic risk monitoring and prevention[14].

Laboratory experiments provide a unique opportunity to control the local fault composition and loading, and thus to understand the local mechanisms responsible for large-scale phenomena. In this study, we consider the case of a laboratory frictional interface along which slow slip is induced by the presence of a heterogeneity in the interface composition. Interfacial heterogeneities affect the macroscopic dynamics of a frictional system[15,16], for example by promoting confined ruptures and thus increasing the frequency of occurrence of rapid slip events[17-19]. Fault inhomogeneities have, moreover, been shown to be prone to exhibit slow slip[20,21]. Our experiment demonstrates a case where slow slip is responsible for the initiation of rapid slip events by acting as a seed crack, which eventually propagates and destabilizes the interface. We highlight the interaction between a decoupled, slowly-slipping zone and coupled, seismically ruptured zones, and the role of this interaction in modifying the stick-slip cycle.

*Corresponding author.
  *Email address:* elsa.bayart@ens-lyon.fr

## Laboratory-fault experiment

A quasi-1D frictional interface is formed by two macroscopically flat solid surfaces in contact, with a granular material embedded over 20% of the total length (Fig. 1a). The system consists of two poly(methyl methacrylate) (PMMA) blocks machined as follows: the contacting surfaces have a 1 μm r.m.s. roughness, while a semi-elliptical shape is hollowed out in the center of each surface, forming an eye-shaped hole when the blocks are in contact, into which grains are inserted. The embedded granular material is made 2D by inserting nylon cylinders of length $l_{cyl} = e$, with $e$ the blocks' thickness, parallel to the z-axis. Different cylinder diameters (from 0.4 to 1.3 mm) are used to prevent crystallization. The cylinders are held in place in the eye-shaped hole by friction and jamming. To force interfacial slip within the granular pile, additional cylinders are fixed to the solid surfaces forming the hole. The granular packing fraction is controlled by the number of cylinders inserted into the hole. PMMA has a strain rate dependent Young's modulus $3 < E < 5.6$ GPa, Poisson ration $\nu = 0.3$, and a Rayleigh wave speed $c_R = 1237 \pm 10$ m/s (plane stress). For nylon, $E = 1.4$ GPa and $\nu = 0.4$.

The blocks are pressed together with a normal displacement resulting in a normal force $F_N \sim 3000\ N$, unless otherwise specified. The shear force $F_S$ is uniformly applied by translating the bottom block at 20 μm/s. $F_N$ and $F_S$ are measured at 315 Hz via load cells. The 2D strain tensor $\varepsilon_{ij}(t)$ is measured at 10 locations, a few millimeters above the interface, and both continuously recorded at 315 Hz and at 4 MHz, when triggered by a rapid event (Methods). The interfacial slip is measured using image correlation of the patterned cylinder faces, imaged at 100 fps, resulting in 8 μm resolution in displacement. Tracking is performed on the cylinders embedded in the eye-shaped hole boundaries and on similar patterns drawn on the block faces above the interface solid-solid sections, allowing slip measurements under the same conditions as for the granular section (Methods).

## Slow slip and stick-slip cycle

We perform reference solid-solid experiments, where the hole is empty, and granular experiments, where cylinders fill the hole. In both cases, the frictional system experiences stick-slip motion (Fig. 1c). The stick-slip frequency is, however, increased in presence of the granular material compared to the reference experiment. The increase is especially noticeable when the density of the granular medium is increased, i.e. an increased number of inserted cylinders (Fig. 1b,c). We compute the normal stress $\sigma_{yy}$ using the strain gages measurements, at 9 locations above the solid-solid sections of the interface and one above the granular section. As $F_N$ is the same for all experiments, $\sigma_{yy}^{gran}$ (resp. $\sigma_{yy}^{solid}$) carried by the granular (resp. solid-solid) section

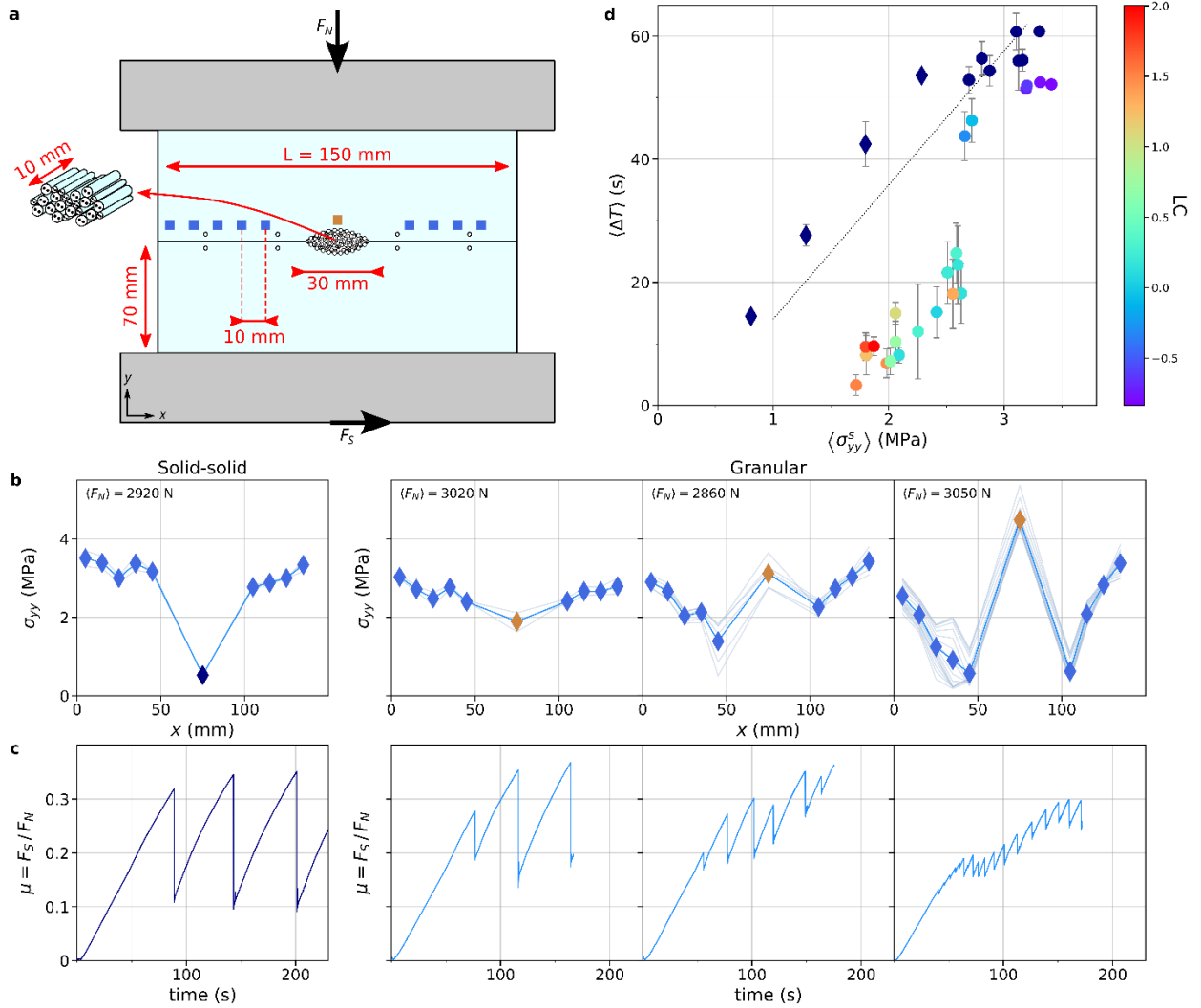

**Fig. 1: Experimental setup and frictional behavior.** (a) Two blocks of PMMA forming a frictional interface with a central eye-shaped hole. The blocks and the hole are (L×H×e) 150×90×10 mm and 30×6×10 mm respectively. Shear force $F_S$ is applied homogeneously by translating the bottom block, while the top block is fixed. Nylon cylinders of diameters 0.4, 0.7, 0.9 and 1.3 mm in approximate volume ratio of 5, 10, 35 and 50% are filling the hole and cylinders of diameter 1.3 mm are inserted into the hole surfaces to roughen it. An array of 10 strain gages measures the 3 components of the 2D strain tensor from 2 to 6 mm above the frictional interface. Axis $x, y, z$ are respectively the shear loading, normal loading and sample thickness directions. (b) Normal stress distributions $\sigma_{yy}(x_{SG})$ along the interface for (left) a solid-solid reference experiment (empty hole) and (right) three granular experiments (hole filled with cylinders), conducted at $F_N \sim$ 3000N. For the granular experiments, from left to right, the density of the granular medium is increased leading to an increased normal stress above the granular section. Light blue diamonds indicate measurements above the solid-solid sections, dark blue above the empty hole and brown above the granular section. (c) Temporal evolution of the friction coefficient $\mu$ for the same experiments as in (b), demonstrating a stick-slip dynamics. (d) Stick-slip period as a function of the average normal stress carried by the solid-solid sections $<\sigma_{yy}^{solid}>$ for solid-solid experiments performed under varying normal force, 750 N$< F_N <$3000 N (dark blue diamonds), reference solid-solid experiments performed at $F_N \sim$3000 N (dark blue circles) and granular experiments performed at $F_N \sim$3000 N but with a varying loading contrast $LC$ (colored circles). The colorbar codes for the loading contrast value.

$$LC = \frac{\sigma_{yy}^{gran} - <\sigma_{yy}^{solid}>}{<\sigma_{yy}>}$$

where $<\sigma_{yy}^{solid}>$ and $<\sigma_{yy}>$ are respectively the averaged normal stresses carried by the solid-solid sections and by the total interface. For a solid-solid experiment, a reduced normal load leads to smaller shear force drops and therefore to an increased stick-slip frequency. We compare the stick-slip period of solid-solid experiments under varying normal force, $750\,N < F_N < 3000\,N$, with granular experiments under $F_N = 3000\,N$ with a varying density, hence varying $<\sigma_{yy}^{solid}>$ and $LC$ (Fig. 1d). The frequency increase in the presence of the granular section is greater than that observed for a corresponding normal load reduction in solid-solid experiments, demonstrating an effect of the compositional heterogeneity on the stick-slip cycle beyond unloading. In the following, we analyze the underlying mechanisms for this intriguing phenomenon, where the presence of a granular material induces an increase in the stick-slip frequency exhibited by the frictional system.

Interfacial slip measurements are used to determine the total slip of the solid-solid and granular sections of the interface, resp. $\delta_{tot}^{solid}(t)$ and $\delta_{tot}^{gran}(t)$, and the cumulated inter-event slip $\delta_{IE}^{solid}(t)$ and $\delta_{IE}^{gran}(t)$, defined as the slip experienced by the interface *excluding* slip occurring during rapid events (Fig. 2a,b). These quantities are averaged over the measurement locations for solid-solid and granular sections. We define the normalized cumulated inter-event slip $S^{\{l\}}(t)$ as:

$$S^{\{l\}}(t) = \frac{\delta_{IE}^{\{l\}}(t)}{\delta_{tot}^{\{l\}}(t)},$$

where $\{l\}$ refers to solid-solid or granular sections. $S^{\{l\}}(t)$ quantifies the coupling of the interface as it corresponds to the fraction of slip occurring during the inter-event periods up to a time $t$. For reference experiments (without grains) measurements show that the entire interface is locked during the inter-event time, with $S^{\{l\}}(t)$ of the order of few percent for both sections, corresponding to the ideal stick-slip motion where slip occurs only during rapid events (Fig. 2c). On the contrary, for a larger loading contrast, corresponding to granular experiments, the granular patch experiences inter-event slip while the solid-solid sections remain almost locked between rapid slip events ($S^{solid} \ll S^{gran}$ in Fig. 2d). We measure the final values of the inter-event cumulated slip, $S_f^{\{l\}}$, as the average over the last few events of an experiment, as a function of the loading contrast $LC$ (Fig. 3a). $S_f^{gran}$ and $S_f^{solid}$ are found to increase with $LC$, but $S_f^{solid}$ always remains below $S_f^{gran}$. This results indicates that differential interfacial slip, increasing with $LC$, is induced along the interface between the granular and solid-solid sections (Fig. 3b). We note that the mean stick-slip frequency of the experiments, $<f_{ss}>$, increases with the differential inter-event cumulated slip $S_f^{gran} - S_f^{solid}$, suggesting that this quantity is a relevant control parameter for the rapid slip events occurrence. Based on strain measurements, we will show that slow slip is localized in the granular patch but also in surrounding solid-solid zones.

## Expansion of the slipping zone with the normal load

Rapid sliding events experienced by a frictional interface are mediated by the propagation of a rupture that weakens the microcontacts resisting shear. These ruptures have been observed in many experimental systems with analog materials[22-25], rocks[26-28], or with a gouge layer[17,18]. They are classical shear cracks for a homogeneous system of two flat solids in contact[24,29]. Hence, an increased occurrence of slip events, as observed in our experiment, indicates an increased frequency of interfacial rupture nucleation and propagation. We will show that the slowly slipping area expands when loading contrast is increased, leading to an early destabilization of the frictional system via rapid rupture propagation. The granular patch serves as a trigger for rupture nucleation, possibly due to shear stress buildup at the patch boundary caused by differential slip among interface sections.

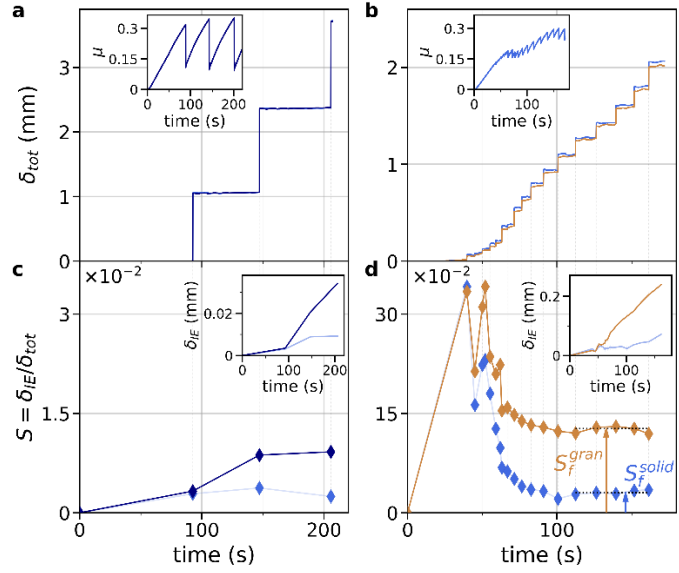

**Fig. 2: Measurements of the interfacial slip.** (a-b) Temporal evolution of the total interfacial slip of the solid-solid sections $\delta_{tot}^{solid}(t)$ (blue line) and of the granular section $\delta_{tot}^{gran}(t)$ (brown line), obtained from the particle tracking method for (a) a solid-solid reference experiment and (b) a granular experiment ($LC = 1.2$). Measurements are averaged over 4 locations for the solid-solid sections and 3 locations for the granular section. (c-d) Temporal evolution of the cumulated inter-event slip $S^{solid}(t)$ (light blue line) and $S^{gran}(t)$ (dark blue line in (c), brown line in (d)), i.e. the interfacial slip excluding the slip occurring during rapid slip events, (c) for the reference experiment and (d) for the granular experiment. Each dot corresponds to one measurement performed at the end of an inter-event period. Inset: the cumulated inter-event slip $\delta_{IE}^{solid}(t)$ and $\delta_{IE}^{gran}(t)$ without normalization (same colors as in main plots).

In Fig. 4a we present the dynamical measurements of the shear strain drop $\Delta\varepsilon_{xy}(x_{SG}, t) = \varepsilon_{xy}(x_{SG}, t) - \varepsilon_{xy}^0(x_{SG})$, where $\varepsilon_{xy}^0(x_{SG})$ is the shear strain before a rupture event at the strain gage locations. The passage of the frictional rupture is revealed by a sudden drop in $\Delta\varepsilon_{xy}(x_{SG}, t)$. Picking up the first variation time for each strain gage provides a measurement of the rupture speed, shown to range from 600 m/s to 2500 m/s, which indicates that both sub-Rayleigh and supershear ruptures occur. In our experiment, each rapid rupture crosses the entire interface. We define the nucleation location of a rupture as the first locations where $\Delta\varepsilon_{xy}(x_{SG}, t)$ starts to depart from 0. Determining this point for all events, we establish histograms of nucleation locations for different ranges of $LC$ (Fig. 4b). For the reference solid-solid experiments, most of the rupture nucleations occurs near one of the patch corners. The distribution is robust when the bottom block is flipped, so the location *is not* imposed by interfacial defects. Fig. 4b clearly shows that the presence of a granular material, even at low density (low $LC$), affects the nucleation location, which then occurs on both sides of the patch. Moreover, as $LC$ increases, the nucleation point is shifted away from the granular patch towards the outer corners of the blocks. Interfacial areas can slip when the frictional microcontacts are already weakened and thus no dynamic rupture can initiate within these areas. Therefore, the fact that rupture nucleation locations move away from the center with the loading contrast indicates that the length of the central slipping zone increases with $LC$. We interpret this result as the fact that a slowly slipping granular section induces the formation of a slipping seed crack, destabilized following a Griffith-type criterion. According to this scenario, the central zone of the interface, including the eye-shaped hole filled with cylinders and adjacent solid-solid zones that do not break via dynamic ruptures, must exhibit slow slip during the inter-event periods. This is indeed what our tracking methods revealed (Fig. 3), but the low spatial resolution of slip measurements prevents a direct measurement of the slipping zone spatial extent. However, low frequency strain measurements allow us to overcome this lack of resolution.

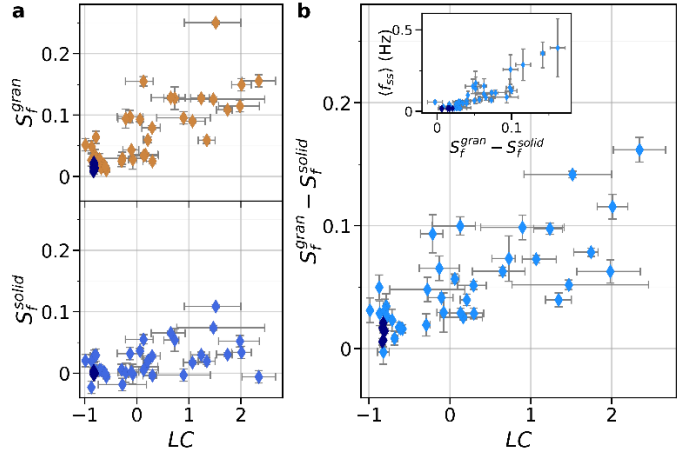

**Fig. 3: Slow slip measurements.** (a) Normalized inter-event cumulated slip (top) $S_f^{gran}$ for the granular section and (bottom) $S_f^{solid}$ for the solid-solid sections of the interface, as a function of the loading contrast $LC$. (b) Differential cumulated slip $S_f^{gran} - S_f^{solid}$ as a function of the loading contrast $LC$. Inset: mean frequency of the stick-slip cycle as a function of $LC$.

For an ideal stick-slip motion, the stick phases correspond to an elastic shear loading, where the shear strain (stress) increases linearly with tangential displacement of the block. In our experiments, this linear loading is indeed observed for some strain gages, while a sublinear evolution of the shear strain is observed for others (Fig. 4c). Since PMMA is an elastic material at low frequency, the sublinear shear strain evolution indicates that interfacial slip takes place. Hence, for each inter-event period, the strain gages exhibiting this sublinear evolution of $\varepsilon_{xy}(x_{SG}, t)$ are detected as a marker of local interfacial slow slip. At each location $x_{SG}$, we count the fraction of nonlinear loading sequences for different ranges of $LC$ (Fig. 4d). Locations with a large number of sublinear evolution of $\varepsilon_{xy}(x_{SG}, t)$ means that slow slip often occurs during the inter-event periods. The larger the central width of the distribution, the longer the slipping zone. Our measurements show that the slipping zones' spatial extent increases with $LC$, which is consistent with the picture obtained from the histograms of nucleation locations (Fig. 4b). A validation of this method is provided by the reference experiments, as in Fig. 4b&d, the nucleation location and the sublinear evolution of the shear strain extend over the same area, from $x = 100$ to $x = 120$ mm.

## Slowly slipping area acting as a nucleation seed

Our combined measurements of the rupture nucleation locations and the evolution of the shear strain during the inter-event periods show that the introduction of a heterogeneity consisting of a granular material induces localized slow slip during the inter-event periods of the stick-slip cycle. As the normal loading applied on the granular section of the interface increases, the slipping zone extends from nearly the size of the patch to almost the entire interface. In parallel, the stick-slip frequency increases with the loading contrast, revealing more frequent rupture initiations for a higher loading of the granular section (Figs. 1&3b).

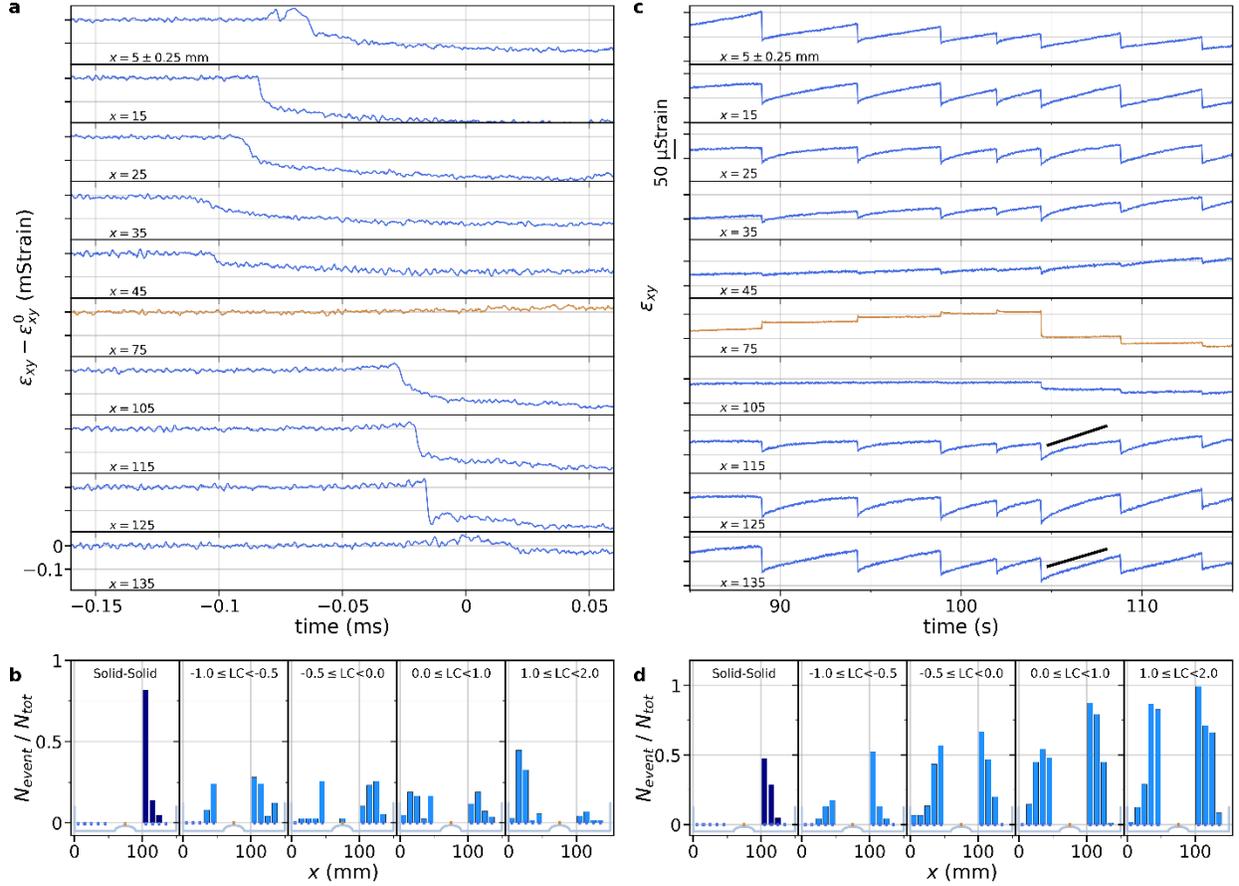

**Fig. 4: Formation of a pre-crack responsible for the early destabilization of the frictional interface.** (a) The shear-strain variation $\varepsilon_{xy}(x_{SG}, t) - \varepsilon_{xy}^0(x_{SG})$ due to the propagation of a rapid rupture, measured at 4 MHz for a granular experiment at $LC = $ -0.6, at the 10 measurement locations (blue lines for strain gages above the solid-solid sections, brown line above the granular patch). (b) Histograms of the rupture initiation locations for (left) the solid-solid reference experiments and (right) the granular experiments for different ranges of $LC$ mentioned on the plots. Each bin corresponds to the ratio of the number of ruptures that have been initiated above a given strain gage. (c) Temporal evolution of the shear strain $\varepsilon_{xy}(x_{SG}, t)$ over a granular experiment at $LC = 1.47$, measured at 315 Hz, for the 10 strain gages. For each event and each strain gage, sublinear evolution of the shear strain during the inter-event period is detected. (d) Counts of the ratio of the number of events exhibiting a sub-linear evolution of the shear strain for each strain gage for (left) the solid-solid reference experiments and (right) the granular experiments, for different ranges of $LC$ mentioned on the plots.

Our interpretation is that the central slipping zone acts as a nucleation center or a seed crack along the interface. In Mode II, a seed crack is an area where the shear stress has been released by interfacial slip. The Griffith criterion expresses the critical stress $\tau_c$ for which a crack of a given length $l$ is destabilized, $\sigma_c = \sqrt{G_c E/(1-\nu^2)l}$, with $G_c$ the interfacial fracture energy. In our experiment, the seed crack length corresponds to the size of the central slipping zone, which increases with $LC$. Hence, the critical stress at which ruptures initiate decreases with $LC$, leading to the triggering of rapid slip events for a lower shear load and therefore, an increased stick-slip frequency.

This mechanism of an extending seed crack leading to an increased stick-slip frequency is not trivial. An obvious scenario would be that the slow slip of the granular patch induces a stress concentration at the junction between the uncoupled zone – the patch – and the partially coupled zone – the solid-solid sections. Therefore, overloading at the patch corners would be responsible for dynamic rupture propagation by reaching the contacts' shear resistance, as observed in different systems[30,31]. However, this is not what we observe. The slow slip within the granular patch induces creep of the neighboring contacts rather than dynamic rupture. Thus, the slowly slipping area affects the rupture dynamics of the entire interface by modifying the rupture nucleation phase, rather than acting locally on the stress distribution. This result is in contrast with other studies where the frictional properties of interfacial heterogeneities are shown to evolve with the sliding history[21,31,32], or where an increased stick-slip frequency is explained by the propagation of ruptures arrested by heterogeneities, which act as a barrier[17-19].

### Extension mechanisms and implications for seismic fault dynamics

What causes microcontacts to creep instead of breaking dynamically? There may be two possible mechanisms involved. The first one is the low rate of the shear overloading induced by the granular slow slip. In fact, frictional contacts age over time, strengthening the interface, and a competition between loading and strengthening may lead to creep rather than dynamic failure[33]. The second one is the shear-induced dilatancy of the granular patch. A dense granular material dilates, i.e. expands, when sheared, which can locally release the normal load on the surrounding areas, leading to a transition to stable sliding[34]. This last mechanism emphasizes the importance of considering the mechanical behavior of heterogeneities in models.

This study highlights an interaction mechanism between an uncoupled, slowly-slipping zone and coupled, locked zones. The boundaries of the uncoupled zone are not restricted to the compositional heterogeneity along the fault, i.e. the granular material, but extends through slow slip or contacts creep of surrounding areas. Is the slow sliding a manifestation of a nucleation front, as observed in experiments[21,30,35] and models[36,37]? The temporal resolution achieved in this experiment is not sufficient to characterize the slow dynamics of contacts detachment and further studies should be devoted to the slow slip region extension mechanism, particularly relevant for the prevention of seismic hazard[8,14].

In summary, we have shown that a slowly slipping area within a frictional interface acts as a seed crack, which induces an early destabilization of the entire interface by lowering the stress level at which the crack propagates, according to a Griffith-like criterion. This mechanism results in a modification of the stick-slip cycle. Our study highlights a fracture mechanics-based description of the effect of a heterogeneity on the frictional interface dynamics. This offers new perspectives for accounting for the complexity of faults in models and for advancing the understanding of the diversity of seismic faults behavior. Finally, our observations could provide a scenario of repeated microearthquakes where slow slip is identified by the triggered seismic events[3].

### Acknowledgements

This work was supported by funding from French National Research Agency (ANR) under Grant No. ANR-20-CE30-0010-01 (project DisRuptInt), from the IDEXLYON Project of the University of Lyon as part of the Investissements d'Avenir Program (ANR-16-IDEX-0005) and from the Fédération de Recherche André-Marie Ampère (FRAMA). The authors thank Mokhtar Adda-Bedia and Cécile Lasserre for helpful discussions and comments.

...## Methods

**Experimental setup.** Experiments were performed using a manual press to apply normal force via a normal displacement, and a motorized stage with speed control (20 µm/s) to apply shear force. The press consists of a rigid steel frame pressing together two rectangular blocks of poly(methyl-methacrylate) (PMMA) (strain rate dependent Young's modulus $3 < E < 5.6$ GPa and Poisson ratio $\nu = 0.3$). The dimensions of the PMMA blocks are 90 mm × 150 mm × 10 mm, clamped along 20 mm, resulting in a free height of 70 mm. The blocks are cut and smoothed on an automated milling machine, then hand-polished (1 µm r.m.s. surface roughness). They are flat to within a 20 µm, checked with a confocal profilometer.

The granular material positioned at the interface consists of nylon cylinders (Young's modulus $E = 1.4$ GPa and Poisson ratio $\nu = 0.4$). It is polydispersed with diameters of 0.4, 0.7, 0.9 and 1.3 mm in approximate volume ratios of respectively 5, 10, 35 and 50%. The cylinders are 10 mm long and are cut from straightened fishing line. Their faces are painted with a pattern that allows optical tracking of their position.

**Optical tracking and slip measurements.** The 3 topmost and bottommost embedded cylinders, positioned at 3.1±0.1 mm above and below the central axis of the interface, are used to track the displacement of the granular interface section. Eight patterns similar to the cylinder faces are painted directly on each block, with their center at 1.4 to 2 mm above the solid-solid sections of the interface, allowing for displacement tracking at 4 locations along this section. The tracking of the cylinders is performed using image correlation on 112 x 1280 pix. images, recorded at 100 fps throughout the entire experiment.

The position recorded for the measurement of the inter-event sliding is calculated by detecting the events, and comparing the average position of the cylinders faces before and after each event, in each section of the interface. The total slip of each section $l$ of the interface $\delta_{tot}^{\{l\}}(t)$ is calculated as:

$$\delta_{tot}^{\{l\}}(t) = (x_{bot}^{\{l\}}(t) - x_{bot}^{\{l\}}(t_0)) - (x_{top}^{\{l\}}(t) - x_{top}^{\{l\}}(t_0))$$

where $x_{top,bot}^{\{l\}}(t)$ is the average position of the top/bottom grains of the corresponding section of the interface and $t_0$ is the initial time of the experiment. To compute $\delta_{IE}^{\{l\}}(t)$ we define the detected time for the $i^{th}$ event $t_i$, $\tau^-$ and $\tau^+$ such that $\delta_{tot}^{\{l\}}(t_i - \tau^-)$ and $\delta_{tot}^{\{l\}}(t_i + \tau^+)$ correspond to the total displacement before and after the $i^{th}$ event, with $\tau^- = 0.05$ s and $\tau^+$ ranging from 0.15 s to 0.5 s, depending on the duration of the inter-event period, to ensure that the system has stopped shaking. We calculate the slip occurring during the inter-event period following event $i$ as $d_i^{\{l\}} = \delta_{tot}^{\{l\}}(t_{i+1} - \tau^-) - \delta_{tot}^{\{l\}}(t_i - \tau^+)$. The cumulated inter-event slip $\delta_{IE}^{\{l\}}(t_i)$ is defined as:

$$\delta_{IE}^{\{l\}}(t_i) = \sum_{m=1}^{i} d_m$$

We define the normalized cumulated inter-event slip $S^{\{l\}}(t_i)$ as:

$$S^{\{l\}}(t_i) = \frac{\delta_{IE}^{\{l\}}(t_i)}{\delta_{tot}^{\{l\}}(t_{i+1} - \tau^-)}$$

The value taken for $S_f^{\{l\}}$ is calculated as the median of the last 5 points of $S^{\{l\}}$ if the experiment includes more than 5 events, or as the median of all the events otherwise. The error bars on $S_f^{\{l\}}$ are calculated as their standard deviation.

**Force, strain and stress measurements.** Forces sensors are embedded in the press frame and allow for low frequency force acquisition at 315 Hz. Strain gages rosettes are installed on one face of the upper solid block, at 1.5 to 2 mm from the interface along the solid-solid sections, and at 6 mm from the central axis of the interface above the granular section, i.e. the elliptical hole. We use 10 rosettes, 4 and 5 above each of the 2 solid-solid sections, 10±0.25 mm apart, and one above the granular section of the interface. They allow for both a continuous acquisition of the loading profile at 315 Hz and a fast burst acquisition of the deformation tensor during a seismic event at 4 MHz. The strain gages are 350 Ω resistances with a 1.84 gage factor, amplified by a factor of 500 using two Anderson loops managing 5 rosettes each[29].

The stresses are calculated from the strains using a plane stress hypothesis, justified by the thinness of the blocks. PMMA is a viscoelastic material, but considering that all our loading phases are quasi-static, we use the low-frequency value of the Young's modulus. An accelerometer, connected to a monostable multivibrator, is used as a trigger to detect the slip events, initiate high speed acquisition of the strain gages and forces signals, and compute the mean frequency for each experiment.

The loading contrast, $LC$, defined for each experiment consists on an average of the loading contrast before each sliding events of the experiment. The errorbars are computed as the first and third quartiles of the $LC$ per event distribution.

The frequency is computed as the mean of $\{1/T_i\}_i$ with $T_i$ the time between event $i$ and $i + 1$. The errorbars are computed as the first and third quartiles of this distribution.

**Correction for the effect of system rotation on interfacial slip measurements.** During the loading phase, due to the relative flexibility of the press frame, rotation occurs between the interface direction and the camera horizontal axis $x_{cam}$ (Supplementary Fig. S1a). It amounts to a maximum of 0.2' per sec, corresponding to up to 0.2° between two events in the solid-solid case. This rotation θ can add a maximum of 20 μm of apparent relative displacement $\delta x$ between the top and bottom cylinders faces. This rotation applies equally to high frequency and low frequency stick-slip cycles and should therefore not modify our conclusions. However, to ensure that it has no influence on the slip measurements, the rotation angle θ(t) between the interface and the camera horizontal axis $x_{cam}$ is calculated at any time, as well as the interface axis $x_{int}(t)$. The position of the cylinders' pattern is then projected on this new axis. To compute the angle θ(t) and $x_{int}(t)$, we compute a linear fit of the top and bottom cylinders patterns relatively to $x_{cam}$, yielding $y_{top}(t) = a_{top}(t) \times x_{cam} + b_{top}(t)$ and $y_{bot}(t) = a_{bot}(t) \times x_{cam} + b_{bot}(t)$. Then $x_{int}(t)$ is calculated as the mean of the two. This solution ensures that the interface is always correctly fitted.

**Elastic deformations at the cylinder's patterns height.** The difference in height between cylinder faces above the granular section and the solid-solid sections, respectively 3.1 and 1.5 to 2 mm, can induce errors in the inter-event sliding computation due to the shear deformation of the blocks (Supplementary Fig. S1b). However, this has been verified to be a negligible correction that does not affect the measurements and conclusions of this study. We evaluate the amplitude of this effect by considering the most unfavorable case, where the shear force attains $F_s^{max} = 1500$ N, corresponding to the maximal shear stress is $\tau_{max} = F_s^{max}/A = 1500/1.5 \times 10^{-3} = 10^6$ Pa, where $A$ is the apparent contact area. The induced displacement $\delta x$ between an upper and a bottom cylinder faces is $\delta x = 2h\tau_{max}/G$, where $h$ is the distance between the cylinders faces and $G = E/2(1 + \nu)$ the shear modulus, yielding to $\delta x^{gran} = 9$ μm and $\delta x^{solid} = 5$ μm. As the values measured for the inter-event slip range from 0 to 300 μm, with a usual 10 μm error bar, this source of error is negligible.